\documentclass[12pt]{article}
\usepackage{amsmath}
\usepackage[dvips]{graphicx}
\usepackage{epsfig}

\newcommand{\bm}{\begin{multiline}}
\newcommand{\beq}{\begin{equation}}
\newcommand{\eeq}{\end{equation}}
\newcommand{\beqs}{\begin{eqnarray}}
\newcommand{\eeqs}{\end{eqnarray}}

\begin{document}

\thispagestyle{empty}

\hfill{}

\hfill{}

\hfill{}

\vspace{32pt}

\begin{center}
\textbf{\Large Charging Black Saturn?}

\vspace{48pt}

\textbf{Brenda Chng,}\footnote{%
E-mail: \texttt{phycmyb@nus.edu.sg}} \textbf{Robert Mann,}\footnote{%
E-mail: \texttt{mann@avatar.uwaterloo.ca}} \textbf{Eugen Radu}\footnote{%
E-mail: \texttt{radu@lmpt.univ-tours.fr}} \textbf{and Cristian Stelea}%
\footnote{%
E-mail: \texttt{stelea@phas.ubc.ca}}

\vspace*{0.2cm}

\textit{$^{1}$Department of Physics, National University of Singapore}\\[0pt]
\textit{2 Science Drive 3, Singapore 117542}\\[.5em]

\textit{$^{2}$Perimeter Institute for Theoretical Physics}\\[0pt]
\textit{31 Caroline St. N. Waterloo, Ontario N2L 2Y5 , Canada}\\[0pt]
\textit{Department of Physics, University of Waterloo}\\[0pt]
\textit{200 University Avenue West, Waterloo, Ontario N2L 3G1, Canada}\\[.5em%
]

\textit{$^{3}$Laboratoire de Math\'ematiques et Physique Th\'eorique}\\[0pt]
\textit{Universit\'e Fran\c{c}ois-Rabelais, Tours, France}\\[.5em]

\textit{$^{4}$Department of Physics and Astronomy, University of British
Columbia}\\[0pt]
\textit{6224 Agricultural Road, Vancouver, BC V6T 1Z1, Canada}\\[.5em]
\end{center}

\vspace{30pt}
\begin{abstract}
We construct new charged  static solutions of the Einstein-Maxwell field equations
in five dimensions via a solution generation technique utilizing the
symmetries of the reduced Lagrangian. By applying our method on the
multi-Reissner-Nordstr\"om solution in four dimensions, we generate the
multi-Reissner-Nordstr\"om solution in five dimensions. We focus on the
five-dimensional solution describing a pair of charged black objects with
general masses and electric charges. This solution includes the double
Reissner-Nordstr\"om solution as well as the charged version of the
five-dimensional static black Saturn. However, all the black Saturn configurations that we found contain either a conical or naked singularity. We also obtain a non-extremal configuration of charged black strings that reduces in the extremal limit to a Majumdar-Papapetrou like solution in five dimensions. 
\end{abstract}
\vspace{32pt}

\setcounter{footnote}{0}

\newpage

\section{Introduction}

Higher dimensional black hole solutions have been known for a long time, for
example the Schwarzschild-Tangherlini black holes, their charged
Reissner-Nordstr\"om cousins, as well as the higher dimensional
generalization of the rotating Kerr solution \cite {Tangherlini:1963bw,Myers:1986un}. In the past few years there has been
remarkable progress in this field, notably the discovery of asymptotically
flat black holes with non-spherical horizon topology. A particularly
interesting case is the five dimensional asymptotically flat black ring
solution whose horizon topology is $S^2\times S^1$ instead of the usual $S^3$
topology of the Schwarzschild solution \cite{Emparan:2001wn}.

The existence of five-dimensional black rings revealed that certain
four-dimensional features of General Relativity cannot be easily extended to
dimensions greater than four. For instance, the celebrated `no-hair' theorem
of four dimensional black hole physics does not hold in more than four dimensions. According to the theorem, an asymptotically flat, stationary
charged black hole is uniquely characterized by its mass, charge and angular
momentum and can only have \textbf{a} horizon with spherical topology. This is
violated in five dimensions where one can have exact solutions describing
black rings with non-spherical horizon topology and at the same time not fully characterized by its conserved charges \cite{Emparan:2004wy,Iguchi:2007is}.

In this paper we are interested in five dimensional multi-black hole
solutions related to the black ring solution. Using the recent
extension of the Weyl formalism to dimensions greater than four \cite{Emparan:2001wk}, the construction of the static five-dimensional
multi-black hole solution was carried out in \cite{Tan:2003jz}. One of
the major tasks in multi-black hole physics is how to maintain the black holes in equilibrium. It turns out that in the static vacuum case in five dimensions, conical singularities are required to generically induce struts of stress energy to counter their mutual gravitational attraction just as in four dimensions. An alternative to conical singularities is to use rotation to keep the black holes apart. This is apparent in the case of a single five-dimensional black ring where its angular momentum provides the necessary force to keep the black ring from collapsing. This mechanism is also present in the asymptotically flat black Saturn solution in five dimensions, where a black hole in the center of a rotating black ring can be in equilibrium if the black ring rotates fast enough \cite{Elvang:2007rd,Evslin:2008py}. One other natural candidate for stabilizing a static black ring is a gauge field, in the simplest case an electromagnetic field, and an exact solution describing an electrically charged static black ring was soon found \cite{Elvang:2003yy,Ida:2003wv,Kunduri:2004da,Yazadjiev:2005hr,Yazadjiev:2005gs,Azuma:2008hs,Yazadjiev:2008pt}.
However the presence of an electric charge alone was found insufficient to
stabilize the black ring and prevent it from collapsing, since conical
singularities in this solution were unavoidable. Nonetheless, by submerging
a charged static black ring into an electric/magnetic background field the
conical singularities were eliminated and the static black ring stabilized. The only drawback of this construction was that, due to the
backreaction of the background electromagnetic field, the black ring was no
longer asymptotically flat.

This leads us to conjecture that by introducing a gauge field to counter the
gravitational forces in a multi-black hole system its constituents could be kept static, with the electrostatic repulsion between two charged objects counteracting their mutual gravitational attraction. For instance, in four dimensions there exist static configurations of extremal Reissner-Nordstr\"om black holes. Similar extremal configurations also exist in higher dimensions \cite{Myers:1986rx}, the supersymmetric configurations in five dimensions have been constructed in \cite{Gauntlett:2004wh,Gauntlett:2004qy} and dynamical solutions exist in lower dimensions \cite{Ohta:2000}. However, the general non-extremal charged multi-black hole solutions are still so far unknown. One may consider a similar situation in the case of a static charged black ring immersed in a background electromagnetic field where the electric field generated by the charged black hole sitting in the center of the static black ring has a stabilizing effect. We thus anticipate the existence of a charged version of the black Saturn in five dimensions, \textit{i.e.} a charged black ring (non-rotating) kept in equilibrium by the electric field of a charged black hole sitting in its center. 
 
In order to verify the conjecture of the existence of a static charged black Saturn in equilibrium, one has to construct the complete multi-black hole solutions in five dimensional Einstein-Maxwell theory. The main purpose of this paper is to show how one accomplishes this goal. However, for simplicity we shall restrict our attention to configurations consisting of only two constituents. These solutions will include as special cases the charged black Saturn  solution, the double non-extremal Reissner-Nordstr\"om solution and the double black string solution, whose extremal limit is precisely a string-like variant of the five dimensional Majumdar-Papapetrou solution. We also note that our generated solution can describe configurations of two black rings (orthogonal or concentric). Although a static black Saturn in $d=5$ Einstein-Maxwell theory has been
constructed in recent works \cite{Yazadjiev:2008ty,Ortaggio:2004kr}, this solution is kept in equilibrium by an external magnetic field and approaches at infinity a Melvin universe background. By contrast, all our generated solutions are asymptotically flat. Unfortunately,  we were unable to find non-singular equilibrium black Saturn configurations: we found that either a conical singularity or a naked curvature singularity must be present. The presence of naked curvature singularities is due to the fact that one `mass' parameter is negative - even though the Komar masses of the constituents are positive. The total ADM mass as measured at infinity is also positive.  

As is well known, Einstein's field equations form a set of nonlinear,
coupled partial differential equations. Solving them analytically by brute
force is a formidable task except in the most simplified cases. However, by
considering spacetime geometries endowed with particular symmetries, it is
sometimes possible to derive solutions in a systematic way. Some of the most
powerful known techniques in constructing exact solutions in General
Relativity in higher dimensions require spacetime geometries to be of the
generalized Weyl class as described in \cite{Emparan:2001wk}. In general, one drawback of the generalized Weyl formalism is that it is limited to $D\leq 5$ since general black holes in $D>5$ dimensions do not
admit $(D-2)$ commuting Killing vectors. For our aim of generating the
general charged multi-black hole solution in five dimensions, this
limitation does not affect us.

The structure of this paper is as follows. We first describe the solution
generating technique that will allow us to lift four-dimensional charged
static configurations to five dimensions. We then use the general double
Reissner-Nordstr\"om solutions in four dimensions as a seed and lift it to
five dimensions. We consider in detail the properties of the charged
double-black hole solution as well as those of the charged black Saturn
configuration. Our solution generating method extends easily to the more
general case of Einstein-Maxwell-Dilaton (EMD) gravity with arbitrary
coupling constant and we derive the charged multi-black hole solutions in
this case. We end with a summary of our work and consider avenues for future
research.


\section{Solution generating technique}

In this section we present a solution generating technique that will map a
general static axisymmetric solution of the Einstein-Maxwell theory in four
dimensions to a five dimensional static axisymmetric solution of the
Einstein-Maxwell-Dilaton (EMD) theory with general dilaton coupling. The
solution generating method will allow us to bypass the actual solving of
Einstein's equations as it is based on a comparison of the reduced
Lagrangians of the two theories in three dimensions and the mapping of the
corresponding scalar fields and electromagnetic potentials. This idea can be
traced back to previous work done in four dimensions to relate stationary
axisymmetric vacuum solutions to solutions of the EMD system \cite%
{Galtsov:1995mb}. However, we show that the analogous mapping in our case
can be further modified by introducing new harmonic functions in the final
solution. This new harmonic `degree of freedom' is essential in the correct
construction of the five dimensional solutions.

Our starting point is the five dimensional Lagrangian describing gravity
coupled to a dilaton field $\phi$ and a $2$-form field strength $F_{(2)}$:
\begin{eqnarray}
\mathcal{L}_{5}=\sqrt{-g}\left[R-\frac{1}{2}(\partial\phi)^2 -\frac{1}{4}%
e^{\alpha\phi}F_{(2)}^2\right],
\end{eqnarray}
where $F_{(2)}=dA_{(1)}$, and the only non-zero component of the $1$-form
gauge potential $A_{(1)}$ is $A_{t}$. We assume that both $A_{t}$ and the
scalar field $\phi$ depend only on the coordinates $\rho$ and $z$.

Let us adopt the following axisymmetric metric ansatz in five dimensions and
assume that $f$, $k$, $l$ and $\mu $ depend on the coordinates $%
\rho $ and $z$ only:
\begin{equation}  \label{blood}
ds_{5}^{2}=-fdt^{2}+ld\varphi ^{2}+kd\chi ^{2}+e^{\mu }(d\rho ^{2}+dz^{2}).
\end{equation}
We now perform a double dimensional reduction down to three dimensions,
first along the coordinate $\chi$ then along the time coordinate $t$. Our metric ansatz is:
\beqs
ds_{5}^2&=&e^{\frac{\phi_1}{\sqrt{3}}}\big[e^{\phi_2}ds_3^2-e^{-\phi_2}dt^2\big]+e^{-\frac{2\phi_1}{\sqrt{3}}}d\chi^2,\nonumber
\eeqs
and one obtains:
\begin{eqnarray}  \label{new4metric1}
ds_{3}^{2} &=&e^{\mu }fk(d\rho ^{2}+dz^{2})+flkd\varphi ^{2} ,  \notag \\
e^{-\phi _{2}} &=&f\sqrt{k} ,~~~~~~~ e^{-\frac{\phi _{1}}{\sqrt{3}}} =\sqrt{k%
} , ~~~~~~~ A_{(1)}=A_{t} dt,
\end{eqnarray}
which is a solution of the equations of motion derived from the following
Lagrangian:
\begin{equation}  \label{5dnotmatter1}
\mathcal{L}_{3}=\sqrt{g}\left[R-\frac{1}{2}(\partial \phi )^{2}-\frac{1}{2}%
(\partial \phi _{1})^{2}-\frac{1}{2}(\partial \phi _{2})^{2}+\frac{1}{2}%
e^{\phi _{2}- \frac{\phi _{1}}{\sqrt{3}}+\alpha \phi }(\partial A_{t})^{2}%
\right].
\end{equation}
We now identify the Lagrangian describing the dynamics of the three
dimensional matter fields as:
\begin{eqnarray}  \label{5dmatter1}
\mathcal{L}_{EMD}^{matter}=\sqrt{g}\left[-\frac{1}{2}(\partial \phi )^{2}-%
\frac{1}{2}(\partial \phi _{1})^{2}-\frac{1}{2}(\partial \phi _{2})^{2}+%
\frac{1}{2}e^{\phi _{2}- \frac{\phi _{1}}{\sqrt{3}}+\alpha \phi }(\partial
A_{t})^{2}\right].
\end{eqnarray}
Consider now the four-dimensional Einstein-Maxwell Lagrangian:
\begin{eqnarray}  \label{4delectric}
\mathcal{L}_4&=&\sqrt{-g}\left[R-\frac{1}{4}\tilde{F}_{(2)}^2\right],
\end{eqnarray}
where $\tilde{F}_{(2)}=d\tilde{A}_{(1)}$ and the only non-zero component of $%
\tilde{A}_{(1)}$ is $\tilde{A}_{t}=\omega$. The solution to the equations of
motion derived from (\ref{4delectric}) is assumed to have the following
static and axisymmetric form:
\begin{eqnarray}  \label{newinitialmetric1}
ds_{4}^{2} &=&-\tilde{f}dt^{2}+\tilde{f}^{-1}\big[e^{2\tilde{\mu}}(d\rho
^{2}+dz^{2})+\rho ^{2}d\varphi ^{2}\big],  \notag \\
{\tilde{A}_{(1)}} &=&\omega dt.
\end{eqnarray}
We next perform a Kaluza-Klein reduction along the timelike direction using the metric ansatz:
\beqs
ds_4^2&=&e^{\psi}ds_3^2-e^{-\psi}dt^2,
\eeqs
 to obtain the following metric and fields in three dimensions:
\begin{eqnarray}  \label{new4metric2}
ds_{3}^{2} &=&e^{2\tilde{\mu}}(d\rho ^{2}+dz^{2})+\rho ^{2}d\varphi ^{2} ,
\notag \\
e^{-\psi } &=&\tilde{f} ,~~~~~~~ \tilde{A}_{(0)t}=\omega,
\end{eqnarray}
where we have denoted the scalar from Kaluza-Klein reduction by $\psi$. The
above is a solution to the equations of motion derived from the three
dimensional Lagrangian:
\begin{equation}
\mathcal{L}_{3}=\sqrt{g}\left[R-\frac{1}{2}(\partial \psi )^{2}+\frac{1}{2}%
e^{\psi }(\partial\omega)^{2}\right],
\end{equation}
The Lagrangian describing the dynamics of the three dimensional matter
fields is:
\begin{eqnarray}  \label{letme}
\mathcal{L}_{EM}^{matter}=\sqrt{g}\left[-\frac{1}{2}(\partial \psi )^{2}+%
\frac{1}{2}e^{\psi }(\partial \omega)^{2}\right].
\end{eqnarray}
In order to relate a solution to the field equations derived from (\ref{letme}) to a solution of the field equations derived from (\ref{5dmatter1}) we shall consider as an intermediary step the following field definitions, starting from a given solution ($\psi,\omega$) of (\ref{letme}):\footnote{Note that with this choice the equations of motion derived from (\ref{5dmatter1}) are indeed satisfied, such that (\ref{letme}) is a consistent truncation of (\ref{5dmatter1}).}
\begin{eqnarray}  \label{terri}
\bar{\phi}&=& \frac{3\alpha}{3\alpha^2+4}\psi,~~~~~~~ \bar{\phi}_1= -{\frac {%
\sqrt {3}}{3\,{\alpha}^{2}+4}}\psi,~~~~~~~ \bar{\phi}_2= \frac{3}{3\alpha^2+4%
}\psi,
\end{eqnarray}
while we also transform the electric $1$-form potential as
\begin{equation}  \label{At-transf}
\bar{A}_{t}=\sqrt{\frac{3}{3\alpha ^{2}+4}}\omega.
\end{equation}
One notices then the following relation between the three dimensional
reduced matter Lagrangians:
\begin{equation}
\mathcal{L}_{EM}^{matter}=\left(\frac{3}{3\alpha ^{2}+4}\right)\bar{%
\mathcal{L}} _{EMD}^{matter}.
\end{equation}
Since we have scaled the reduced three-dimensional matter Lagrangian by a
constant factor, in order to match the solutions of the equations of motion
derived from the above lagrangians we also have to modify their three
dimensional geometries such that the Ricci tensor of the metric (\ref%
{new4metric1}) is essentially a constant rescaling of the Ricci tensor of the
metric (\ref{new4metric2}), the scaling factor being $\frac{3}{3a^2+4}$.
By comparing the three-dimensional geometries and taking into consideration the special properties of the Weyl-Papapetrou ansatz (\ref{new4metric2}) in three dimensions (see \cite{Galtsov:1995mb} for more details) this can be easily accomplished by taking:\footnote{This is the scaling symmetry property used in \cite{Chng:2006gh} to derive new solutions in four dimensions.}
\begin{eqnarray}  \label{breath}
e^{\bar{\mu}}\bar{f}\bar{k}&\equiv&\left(e^{2\tilde{\mu}}\right)^{\frac{3}{%
3\alpha^2+4}}, ~~~~~~~ \bar{f}\bar{k}\bar{l}\equiv\rho^2.
\end{eqnarray}

One can check that the `barred' fields $(\bar{\mu}, \bar{f}, \bar{k}, \bar{l}%
, \bar{A}_t)$ solve the equations of motion derived from (\ref{5dmatter1})
as expected. Note that there also exists a freedom in defining the
scalar fields (\ref{terri}), which can be seen from considering new scalars $%
\phi=\bar{\phi}$, $\phi_{1}=\bar{\phi} _{1}-\sqrt{3}h$ and $\phi_{2}=\bar{%
\phi} _{2}-h$ such that the new matter Lagrangian can be written as:
\begin{eqnarray}  \label{5dmatter2}
\mathcal{L}_{(1)}^{matter}&=&\sqrt{g}\left[-\frac{1}{2}(\partial \phi )^{2}-%
\frac{1}{2}(\partial\phi _{1})^{2}-\frac{1}{2}(\partial\phi _{2})^{2}+\frac{1%
}{2}e^{\phi _{2}-\frac{\phi _{1}}{\sqrt{3}}+\alpha \phi }(\partial A_{t})^{2}%
\right]  \notag \\
&=&\sqrt{g}\left[-\frac{1}{2}(\partial \bar\phi )^{2}-\frac{1}{2}%
(\partial\bar\phi _{1})^{2}-\frac{1}{2}(\partial \bar\phi _{2})^{2}+\frac{1}{%
2}e^{\bar\phi _{2}-\frac{\bar\phi _{1}}{\sqrt{3}}+\alpha \bar\phi }(\partial
\bar{A}_{t})^{2}+4(\partial h)^2\right],
\end{eqnarray}
where in the second line of the above equality we used (\ref{terri}) to eliminate the cross-terms containing products of the `barred' scalar fields with $h$. The field $h$ is thus decoupled from the other matter fields.

Notice that in order to obtain a five dimensional solution described by the
scalar fields $\phi$, $\phi_1$ and $\phi_2$ the initial `barred' five
dimensional Einstein-Maxwell-Dilaton solution should be modified to
accommodate the extra scalar field $h$. Notice further that $h$ must be a harmonic function (as can be seen from its equations of motion) and, moreover, since the `barred' five-dimensional EMD fields are not directly coupled to it, its
gravitational backreaction is easily taken care of by introducing a new
function $\gamma$ such that:
\begin{eqnarray}  \label{gammap1a}
\partial_\rho{\gamma}&=&\rho[(\partial_\rho h)^2-(\partial_z h)^2],~~~~~~~
\partial_z{\gamma}=2\rho(\partial_\rho h)(\partial_z h).
\end{eqnarray}
Hence, given a harmonic function $h$, we can solve (\ref{gammap1a}) for $%
\gamma$, which can then be substituted in the following metric:
\begin{eqnarray}
ds_3^{2}&=&(e^{\bar{\mu}}\bar{f}\bar{k})e^{2\gamma}(d\rho
^{2}+dz^{2})+\rho^2d\varphi ^{2},  \label{terrible}
\end{eqnarray}
to obtain a solution to the modified Lagrangian (\ref{5dmatter2}).

Taking into account the scaling of the three dimensional metric and presence
of the harmonic function $h$, we have the following relations:
\begin{eqnarray}  \label{ya}
e^{\mu}fk&\equiv&\left(e^{2\tilde{\mu}}\right)^{\frac{3}{3\alpha^2+4}%
}e^{2\gamma} ,~~~~~~~ fkl\equiv\rho^2,
\end{eqnarray}
where $\gamma$ can be found from (\ref{gammap1a}) once $h$ is known.

Let us now summarize the results of our solution generating method. One
reads off the functions $\tilde{f}$, $\omega$ and $e^{\tilde{\mu}}$ from the
four-dimensional metric (\ref{newinitialmetric1}) and then substitutes them
into the transformations (\ref{terri}) and (\ref{At-transf}). The harmonic
function $h$ can alter the metric only along the spatial directions and its
form is guessed by imposing a desired background geometry in the
final solution. Using (\ref{new4metric1}) and (\ref{ya}), one then computes $%
f, k, l$ and $e^{\mu}$ in terms of $e^{\tilde{\mu}}$, $\tilde{f}$, $h$ and $%
\gamma$. The result is then a new EMD solution in five dimensions, which can
be written as:
\begin{eqnarray}  \label{new5Dkfl}
ds_{5}^{2}=-\tilde{f}^{\frac{4}{3\alpha^2+4}}dt^{2}+\tilde{f}^{-\frac{2}{%
3\alpha^2+4}}\bigg[e^{2h}d\chi ^{2}+e^{\frac{6\tilde{\mu}}{3\alpha^2+4}%
+2\gamma-2h}(d\rho ^{2}+dz^{2})+\rho^2e^{-2h}d\varphi ^{2}\bigg],
\end{eqnarray}
while the $1$-form potential and the dilaton are given by:
\begin{eqnarray}
A_{(1)}&=&\sqrt{\frac{3}{3\alpha^2+4}}\omega dt,~~~~~~~ e^{-\phi}=\tilde{f}^{%
\frac{3\alpha}{3\alpha^2+4}}.
\end{eqnarray}
Solutions of the pure Einstein-Maxwell theory in five dimensions are simply
obtained from the above formulae by taking $\alpha=0$. In the following
sections we shall focus on this case.

\section{Multi-Reissner-Nordstr\"{o}m solutions in five dimensions}

As a check of the technique presented in the last section, we will first map
the four-dimensional Reissner-Nordstr\"{o}m solution to the five-dimensional
Reissner-Nordstr\"{o}m solution. We then use the four-dimensional
double-Reissner-Nordstr\"om solution in a form recently given by Manko \cite%
{Manko:2007hi} as the seed to generate the double-Reissner-Nordstr\"om solution in five dimensions. This four-dimensional solution has been recently
re-derived in \cite{Alekseev:2007re} by using a monodromy transform approach.


\subsection{Single Reissner-Nordstr\"om black holes and charged black rings
in five dimensions}


The four-dimensional Reissner-Nordstr\"{o}m solution is written in Weyl form
as \cite{Emparan:2001bb}:
\begin{eqnarray}  \label{RN4dim}
ds^2&=&-\tilde{f}dt^2+\tilde{f}^{-1} \big[e^{2\tilde{\mu}}(d\rho^2+dz^2)+%
\rho^2 d\varphi^2\big], \\
\omega&=&-\frac{4q}{r_1+r_2+2m},~~~~~~~\tilde{f}=\frac{(r_{1}+r_{2})^2-4%
\sigma^2}{(r_{1}+r_{2}+2m)^2},~~~~~~~ e^{2\tilde{\mu}}=\frac{%
(r_{1}+r_{2})^2-4\sigma^2}{4r_{1}r_{2}},  \notag
\end{eqnarray}
where
\begin{eqnarray}
r_{1}&=&\sqrt{\rho^2+(z-\sigma)^2}, ~~~~~~~ r_{2}=\sqrt{\rho^2+(z+\sigma)^2}.
\end{eqnarray}
Note that $\sigma=\sqrt{m^2-q^2}$ and $m$ denotes the mass and $q$ the
charge.

The five-dimensional Reissner-Nordstr\"om metric is given by (\ref{new5Dkfl}%
) with $\alpha=0$ once we use a suitable harmonic function $h$ to ensure
that the generated five-dimensional metric is also asymptotically flat. With
hindsight, we find that the appropriate $h$ is given by:
\begin{eqnarray}
e^{2h}&=&(r_{2}+(z+\sigma))\left(\frac{r_{1}+(z-\sigma)} {r_{2}+(z+\sigma)}%
\right)^{\frac{1}{2}}\cr &=&\left[(r_{2}+\zeta_2)(r_{1}+\zeta_1)\right]^{%
\frac{1}{2}},
\end{eqnarray}
and we can now find $\gamma$ from (\ref{gammap1a}):
\begin{eqnarray}
e^{2\gamma}&=&\frac{[(r_{2}+\zeta_2)(r_{1}+\zeta_1)]^{\frac{1}{2}}} {%
[8r_{1}r_{2}Y_{12}]^{\frac{1}{4}}}.
\end{eqnarray}
where $\zeta_1=z-\sigma$, $\zeta_2=z+\sigma$ and $Y_{12}=r_1r_2+\zeta_1%
\zeta_2+\rho^2$. The first factor in $e^{2h}$ moves the semi-infinite rod $z<-\sigma$
from the $\varphi$ direction to the $\chi$ direction, while the second
factor corresponds to a `correction' of the black hole horizon. It turns
out that we will have to take such horizon corrections into account for each
horizon when describing multi-black objects in five dimensions, while the
rod-moving terms in $h$ can be read from the expected rod structure in the final geometry.

We thus obtain:\footnote{%
Note that $2Y_{12}=(r_1+r_2)^2-4\sigma^2$.}
\begin{eqnarray}
ds_{5}^2&=&-\frac{(r_{1}+r_{2})^2-4\sigma^2}{(r_{1}+r_{2}+2m)^2}dt^2+\frac{%
r_1+r_2+2m}{\sqrt{2Y_{12}}}\bigg[\sqrt{(r_{2}+\zeta_2)(r_{1}+\zeta_1)}%
d\chi^2+\frac{\sqrt{2Y_{12}}}{4r_1r_2}(d\rho^2+dz^2)  \notag \\
&&+\frac{\rho^2d\varphi^2}{\sqrt{(r_{2}+\zeta_2)(r_{1}+\zeta_1)}}\bigg].
\end{eqnarray}

Let us now convert the above metric from cylindrical coordinates $(\rho, z)$ to polar
coordinates $(r,\theta)$ by using the relations \cite{Emparan:2001wk}:
\begin{eqnarray}
\rho^2&=&r^2(r^2-4\sigma)\sin^2\theta\cos^2\theta,~~~~~~~ z=\frac{1}{2}%
(r^2-2\sigma)\cos2\theta.  \label{coordtransfsingleRN}
\end{eqnarray}
We obtain:
\begin{eqnarray}
ds_{5}^2&=&-\frac{r^2(r^2-4\sigma)}{(r^2+2(m-\sigma))^2}dt^2 +\frac{%
r^2+2(m-\sigma)}{r^2}\Big(\frac{r^2}{r^2-4\sigma}dr^2+r^2(d\theta^2+\sin^2%
\theta d\varphi^2+\cos^2\theta d\chi^2)\Big)  \notag \\
&&  \notag \\
&=&-H^{-2}(r)f(r)dt^2+H(r)\left(f(r)^{-1}dr^2+r^2d\Omega^2_3%
\right),~~~~~A_t=-\frac{2\sqrt{3}\sqrt{m^2-\sigma^2}}{r^2+2(m-\sigma)},
\end{eqnarray}
where
\begin{eqnarray}
H(r)&=&1+\frac{2(m-\sigma)}{r^2},~~~~~~~ f(r)=1-\frac{4\sigma}{r^2},
\end{eqnarray}
which is indeed the five-dimensional Reissner-Nordstr\"{o}m solution. It
should be clear that if we relax the $\alpha=0$ condition one can obtain from (%
\ref{new5Dkfl}) the dilatonic black hole found previously in \cite%
{Gibbons:1987ps}.

If one chooses the following harmonic function $h$ instead:
\begin{eqnarray}
e^{2h}&=&(r_{0}+(z+\sigma_0))\left(\frac{r_{1}+(z-\sigma)} {r_{2}+(z+\sigma)}%
\right)^{\frac{1}{2}},\nonumber\\
&=&(r_{0}+\zeta_0)\left(\frac{r_{1}+\zeta_1} {r_{2}+\zeta_2}
\right)^{\frac{1}{2}},
\end{eqnarray}
one readily sees that the rod structure of the final solution corresponds to
a static black ring. Here we denote $r_0=\sqrt{\rho^2+(z+\sigma_0)^2}$ and $\zeta_0=z+\sigma_0$,
where $\sigma_0>\sigma>0$. We can now find $\gamma$ from (\ref{gammap1a}):
\begin{eqnarray}
e^{2\gamma-2h}&=&\frac{1}{K_0r_0}\left(\frac{Y_{02}}{Y_{01}}\right)^{\frac{1%
}{2}}\left(\frac{4Y_{12}}{r_1r_2}\right)^{\frac{1}{4}},
\end{eqnarray}
where $Y_{ij}=r_ir_j+\zeta_i\zeta_j+\rho^2$, $%
i,j=0,1,2$ and $K_0$ is an integration constant.  In Weyl coordinates the charged black ring solution is then found to be:
\beqs
ds^2&=&-\frac{(r_{1}+r_{2})^2-4\sigma^2}{(r_{1}+r_{2}+2m)^2}dt^2+\frac{r_1+r_2+2m}{\sqrt{2Y_{12}}}\bigg[(r_0+\zeta_0)\sqrt{\frac{r_{1}+\zeta_1}{r_{2}+\zeta_2}}d\chi^2+\frac{2Y_{12}}{K_0r_0r_1r_2}\sqrt{\frac{Y_{02}}{Y_{01}}}(d\rho^2+dz^2)\nonumber\\
&&+\frac{\rho^2d\varphi^2}{r_0+\zeta_0}\sqrt{\frac{r_2+\zeta_2}{r_1+\zeta_1}}\bigg],~~~~~~~
A_t=-\frac{\sqrt{3}}{2}\frac{4\sqrt{m^2-\sigma^2}}{r_1+r_2+2m}.
\eeqs
The metric of the uncharged static black ring (see for instance equations $(4.15-4.18)$ in \cite{Emparan:2001wk}) is recovered in the limit $m=\sigma$, thus confirming that the above solution describes a static black ring in Weyl coordinates. We have thus generated the static charged black ring as a solution of
Einstein-Maxwell-Dilaton system in five dimensions, a solution previously
found in \cite{Kunduri:2004da}.


\subsection{The double-Reissner-Nordstr\"om solution in five dimensions}

We start from the four-dimensional double Reissner-Nordstr\"{o}m solution in
the parameterization given recently by Manko in \cite{Manko:2007hi}. In our
notation, the four-dimensional fields read:
\begin{equation}
\tilde{f}=\frac{A^{2}-B^{2}+C^{2}}{(A+B)^{2}},~~~~~~e^{2\tilde{\mu}}=\frac{%
A^{2}-B^{2}+C^{2}}{16\sigma _{1}^{2}\sigma _{2}^{2}(\nu
+2k)^{2}r_{1}r_{2}r_{3}r_{4}},~~~~~~~\omega =-\frac{2C}{A+B},  \label{Manko}
\end{equation}%
where:
\begin{eqnarray}
A &=&\sigma _{1}\sigma _{2}[\nu
(r_{1}+r_{2})(r_{3}+r_{4})+4k(r_{1}r_{2}+r_{3}r_{4})]-(\mu ^{2}\nu
-2k^{2})(r_{1}-r_{2})(r_{3}-r_{4}),  \notag \\
B &=&2\sigma _{1}\sigma _{2}[(\nu M_{1}+2kM_{2})(r_{1}+r_{2})+(\nu
M_{2}+2kM_{1})(r_{3}+r_{4})]  \notag \\
&&-2\sigma _{1}[\nu \mu (Q_{2}+\mu )+2k(RM_{2}+\mu Q_{1}-\mu
^{2})](r_{1}-r_{2})  \notag \\
&&-2\sigma _{2}[\nu \mu (Q_{1}-\mu )-2k(RM_{1}-\mu Q_{2}-\mu
^{2})](r_{3}-r_{4}),  \notag \\
C &=&2\sigma _{1}\sigma _{2}\{[\nu (Q_{1}-\mu )+2k(Q_{2}+\mu
)](r_{1}+r_{2})+[\nu (Q_{2}+\mu )+2k(Q_{1}-\mu )](r_{3}+r_{4})\}  \notag \\
&&-2\sigma _{1}[\mu \nu M_{2}+2k(\mu M_{1}+RQ_{2}+\mu R)](r_{1}-r_{2})
\notag \\
&&-2\sigma _{2}[\mu \nu M_{1}+2k(\mu M_{2}-RQ_{1}+\mu R)](r_{3}-r_{4}),
\end{eqnarray}%
with constants:
\begin{eqnarray}
\nu &=&R^{2}-\sigma _{1}^{2}-\sigma _{2}^{2}+2\mu
^{2},~~~~~~~k=M_{1}M_{2}-(Q_{1}-\mu )(Q_{2}+\mu ),  \notag \\
\sigma _{1}^{2} &=&M_{1}^{2}-Q_{1}^{2}+2\mu Q_{1},~~~~~~~\sigma
_{2}^{2}=M_{2}^{2}-Q_{2}^{2}-2\mu Q_{2},~~~~~~~\mu =\frac{%
M_{2}Q_{1}-M_{1}Q_{2}}{M_{1}+M_{2}+R},
\end{eqnarray}%
while $r_{i}=\sqrt{\rho ^{2}+\zeta _{i}^{2}}$, for $i=1..4$, with:
\begin{equation}
\zeta _{1}=z-\frac{R}{2}-\sigma _{2},~~~~~\zeta _{2}=z-\frac{R}{2}+\sigma
_{2},~~~~~\zeta _{3}=z+\frac{R}{2}-\sigma _{1},~~~~~\zeta _{4}=z+\frac{R}{2}%
+\sigma _{1}.
\label{ai}
\end{equation}%
This solution is parameterized by five independent parameters and describes
the superposition of two general Reissner-Nordstr\"{o}m black holes, with
masses $M_{1,2}$, charges $Q_{1,2}$ and $R$ the coordinate distance
separating them. For a detailed discussion of its properties we refer the
reader to \cite{Manko:2007hi} and the references therein. We shall note here
that in general the function $e^{2\tilde{\mu}}$ can be determined up to a
constant and its precise numerical value has been fixed here by allowing the
presence of conical singularities only in the portion in between the black
holes along the $\varphi $ axis. Consequently one has:
\begin{equation}
e^{2\tilde{\mu}}|_{\rho =0}=\left( \frac{\nu -2k}{\nu +2k}\right) ^{2},
\label{strutManko}
\end{equation}%
for $-R/2+\sigma _{1}<z<R/2-\sigma _{2}$ and $e^{2\tilde{\mu}}|_{\rho =0}=1$
elsewhere.

Using the results from the previous section, the corresponding
five-dimensional solution of the Einstein-Maxwell system reads:
\begin{eqnarray}  \label{rel1}
ds_{5}^2&=&-\tilde{f}dt^2+\tilde{f}^{-\frac{1}{2}}\bigg[e^{2h}d\chi^2+e^{-2h}%
\big[e^{3\tilde{\mu}/2+2\gamma}(d\rho^2+dz^2)+\rho^2d\varphi^2\big]\bigg],
\notag \\
A_t&=&-\frac{\sqrt{3}C}{A+B}.  \label{final5}
\end{eqnarray}
So far the harmonic function $h$ is still arbitrary. One can see that $h$'s
presence can alter the rod structure of the final solution along the $\chi$
and $\varphi$ directions. By carefully choosing the form of $h$, we can construct the appropriate rod structures to describe configurations involving black holes, black rings, or a combination of black holes and black rings. Once we pick a suitable $h$, $\gamma$ is easily found by integrating (\ref{gammap1a}). Let us illustrate this by considering three important cases.


\subsubsection{Double-Reissner Nordstr\"om black holes}


To describe a configuration of two black holes, it turns out that the
appropriate choice for the harmonic function $h$ is:
\begin{eqnarray}
e^{2h}&=&\frac{\sqrt{(r_1+\zeta_1)(r_2+\zeta_2)(r_3+\zeta_3)(r_4+\zeta_4)}}{%
r_0+\zeta_0},
\end{eqnarray}
where we denote $r_i=\sqrt{\rho^2+\zeta_i^2}$ and $\zeta_i=z-a_i$ for $i=0 ..
4 $, $a_i$ can be read from (\ref{ai}) and $a_0=0$. The corresponding
rod structure of this solution is given in Figure $1a)$.
\begin{figure}[tbp]
\par
\begin{center}
\includegraphics{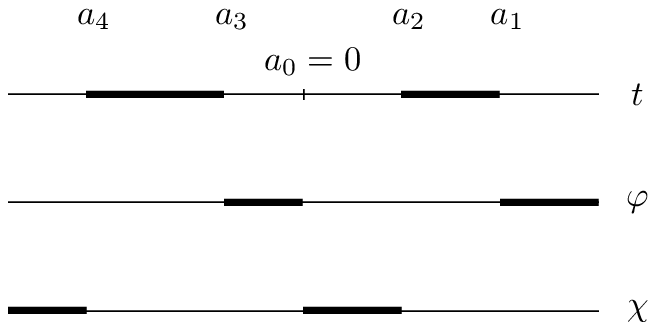} \hskip1.5cm \includegraphics{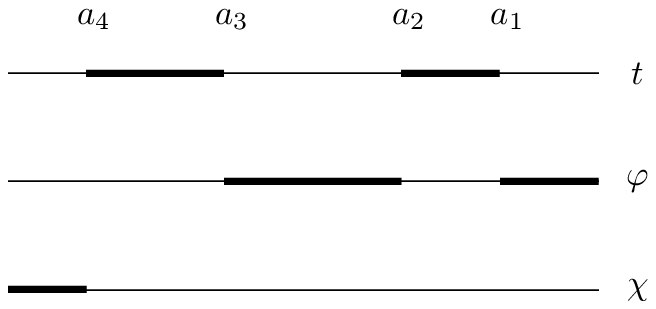}
\end{center}
\caption{Rod structures of (a) the double-black hole system, and (b) the
black Saturn.}
\label{rod}
\end{figure}
Utilizing methods from \cite{Emparan:2001wk} one can easily integrate (\ref%
{gammap1a}) to find:
\begin{eqnarray}
e^{2\gamma-2h}&=&\frac{1}{K_0r_0}\frac{\left(Y_{01}Y_{02}Y_{03}Y_{04}%
\right)^{\frac{1}{2}}}{%
\left(r_1r_2r_3r_4Y_{12}Y_{13}Y_{14}Y_{23}Y_{24}Y_{34}\right)^{\frac{1}{4}}}
\end{eqnarray}
Here $K_0$ is an arbitrary constant whose value will be fixed later on,
and $Y_{ij}=r_ir_j+\zeta_i\zeta_j+\rho^2$. The final solution
solution describing the general double-Reissner-Nordstr\"om black hole is
then given by (\ref{final5}) for this particular choice of the harmonic function $h$. The solution depends on five parameters and corresponds
physically to the masses, charges of the two black holes and the
distance between them.

Before we embark on a discussion of its physical properties, let us first consider the structure of the conical singularities along the axis. To define a
conical singularity for a rotational axis with angle $\theta$ one computes
the proper circumference $C$ around the axis and its proper radius $R$ and
define:
\begin{eqnarray}
\alpha&=&\frac{dC}{dR}|_{R=0}=\lim_{\rho\rightarrow 0}\frac{\sqrt{%
g_{\theta\theta}}\Delta\theta}{\int_0^{\rho}\sqrt{g_{\rho\rho}}d\rho}%
=\lim_{\rho\rightarrow 0}\frac{\partial_{\rho}\sqrt{g_{\theta\theta}}%
\Delta\theta}{\sqrt{g_{\rho\rho}}},
\end{eqnarray}
where $\Delta\theta$ is the period of $\theta$. The presence of a conical
singularity is now expressed by means of:
\begin{eqnarray}
\delta&=&2\pi-\alpha,
\end{eqnarray}
such that $\delta>0$ corresponds to a conical deficit (a `cosmic string'),
while $\delta<0$ corresponds to a conical excess (a `strut').

Consider now the $ds_{\chi \rho }^{2}$ part of the metric. It turns out that
conical singularities cannot be avoided and must be present either along $%
0<z<R/2-\sigma _{2}$ where:
\begin{equation*}
\delta _{\chi }=2\pi -\Delta \chi \sqrt{\frac{K_{0}}{\sqrt{8}}}\left( \frac{%
\nu +2k}{\nu -2k}\right) ^{\frac{3}{4}}\left( \frac{16\big[(R+\sigma
_{2})^{2}-\sigma _{1}^{2}\big]\big[(R-\sigma _{2})^{2}-\sigma _{1}^{2}\big]}{%
(R^{2}-4\sigma _{2}^{2})^{2}}\right) ^{\frac{1}{4}},
\end{equation*}%
or along $z<-R/2-\sigma _{1}$, where we find:
\begin{equation*}
\delta _{\chi }=2\pi -\Delta \chi \sqrt{\frac{K_{0}}{\sqrt{8}}}.
\end{equation*}

If we set the period of $\chi $ be $\Delta \chi =2\pi $ and choose to have a
regular outer axis (for $z<-R/2-\sigma _{1}$) we must set $K_{0}=\sqrt{8}$.
Similarly, for the $ds_{\varphi \rho }^{2}$ part of the metric, it turns out
that conical singularities cannot be avoided and must be present either
along $-R/2+\sigma _{1}<z<0$, where:
\begin{equation*}
\delta _{\varphi }=2\pi -\Delta \varphi \sqrt{\frac{K_{0}}{\sqrt{8}}}\left(
\frac{\nu +2k}{\nu -2k}\right) ^{\frac{3}{4}}\left( \frac{16\big[(R+\sigma
_{2})^{2}-\sigma _{1}^{2}\big]\big[(R-\sigma _{2})^{2}-\sigma _{1}^{2}\big]}{%
(R^{2}-4\sigma _{1}^{2})^{2}}\right) ^{\frac{1}{4}},
\end{equation*}%
or along $R/2+\sigma _{2}<z$, where we find:
\begin{equation*}
\delta _{\varphi }=2\pi -\Delta \varphi \sqrt{\frac{K_{0}}{\sqrt{8}}}.
\end{equation*}%
Again, to have a regular outer axis we set $\Delta \varphi =2\pi $ and $%
K_{0}=\sqrt{8}$, in which case we still have a conical singularity on the
axis in between the black holes. Therefore, if one demands the metric to be
asymptotically flat with a regular outer axis, there will be conical defects
between the black holes. Their presence is physically expected since our
solution is static and therefore the conical defects should correspond to forces balancing the gravitational and electromagnetic forces in
between the black holes.

One might wonder if there are some values of the parameters characterizing
the solution for which the conical defects vanish. Our numerical investigation of this issue seems to imply a negative answer (the numerical methods used here are similar to those described in
the black Saturn case). Although further work is clearly necessary, it appears that, similar to the
four-dimensional case, equilibrium configurations require the presence of
some unphysical features of the constituent black holes \cite%
{Manko:2007hi,Alekseev:2007re,Perry:1996ja,Alekseev:2007gt}.

Let us consider now some special limits of the above solution. First, in
order to prove that this solution describes two Reissner-Nordstr\"om black
holes, note that one can recover the individual black hole metric by pushing
the other black hole to infinity. For example, to recover the metric for the
second black hole (described by the parameters $M_2$ and $Q_2$) one has to
first shift the $z$-coordinate $z\rightarrow z-R/2$ (\textit{i.e.} positioning ones center on its horizon's rod) \textbf{and then} take the infinite separation limit $R\rightarrow\infty$.
From the general expressions in (\ref{Manko}), one notes that in this limit $%
\nu\sim R^2$, $\mu\sim 0$, $\sigma_i=\sqrt{M_i^2-Q_i^2}$, for $i=1..2$, $%
k=M_1M_2-Q_1Q_2$ and:
\begin{eqnarray}
A\sim2\sigma_1\sigma_2R^3(r_1+r_2),~~~~~B\sim4\sigma_1\sigma_2R^3M_2,~~~~~C%
\sim4\sigma_1\sigma_2R^3Q_2,
\label{pushaway}
\end{eqnarray}
Therefore the general solution (\ref{Manko}) reduces to (\ref{RN4dim}).
Also, by taking this limit in the harmonic function $h$ one obtains:
\begin{eqnarray}
e^{2h}&=&2\left[(r_{2}+\zeta_2)(r_{1}+\zeta_1)\right]^{\frac{1}{2}},~~~~~~~
e^{2\gamma-2h}=\frac{1} {[64r_{1}r_{2}Y_{12}]^{\frac{1}{4}}}.
\end{eqnarray}
Gathering all these results together and performing the coordinate
transformation (\ref{coordtransfsingleRN}) one readily checks that the
solution indeed reduces to the five-dimensional Reissner-Nordstr\"om black
hole with conical singularities attached in the $\chi$ direction.
Similarly, if one centers on the black hole on the left and pushes the other black hole to infinity one obtains the metric of a single Reissner-Nordstr\"om black hole with a conical singularity attached along
the $\varphi$ direction. As we shall see below, these conical singularities
are unavoidable as they are inherited from the background geometry.

The uncharged case corresponds to setting $Q_1=Q_2=0$ and noting that the
four-dimensional seed solution (\ref{Manko}) reduces in this case to the
Israel-Khan solution \cite{IsraelKhan} describing two neutral black holes,
with the above choice of the harmonic function $h$ one readily checks that
one obtains the five-dimensional uncharged double-black hole
solution constructed in \cite{Tan:2003jz}.

The extremal charged limit of the above solution corresponds to taking the
limits $M_1=Q_1$ and $M_2=Q_2$. This leads to $\sigma_1=\sigma_2=k=\mu=0$
and, in consequence, $r_1=r_2$ and $r_3=r_4$. This extremality limit must be
taken with care as one finds that the quantities $A$, $B$, $C$ all vanish in
this case. However, from the general expression of the metric functions in (%
\ref{Manko}) one can nonetheless find the metric functions in the extremal
case to be:
\begin{eqnarray}
\tilde{f}_e&=&\left(1+\frac{M_1}{r_3}+\frac{M_2}{r_1}\right)^{-2},~~~~~~~
e^{2\tilde{\mu}}|_{e}=1.  \label{extremal4dManko}
\end{eqnarray}
Using these expressions in (\ref{final5}) one finally obtains the charged
double-black hole solution previously found in \cite{Tan:2003jz}:
\begin{eqnarray}
ds_{ext}^2&=&-H(r)^{-2}dt^2+H(r)\bigg[\frac{(r_1+\zeta_1)(r_3+\zeta_3)}{%
r_0+\zeta_0}d\chi^2+\frac{Y_{01}Y_{03}}{4r_0r_1r_3Y_{13}}(d\rho^2+dz^2)+%
\frac{(r_0+\zeta_0)\rho^2d\varphi^2}{(r_1+\zeta_1)(r_3+\zeta_3)}\bigg],
\notag \\
A_t&=&-\frac{\sqrt{3}}{2}H(r)^{-1},~~~~~~H(r)=1+\frac{M_1}{r_3}+\frac{M_2}{%
r_1}.  \label{multiedward}
\end{eqnarray}
Moreover, in absence of the black holes (we set $M_1=M_2=0$) the background
geometry is found to be:
\begin{eqnarray}
ds_{bkg}^2&=&-dt^2+\frac{(r_1+\zeta_1)(r_3+\zeta_3)}{r_0+\zeta_0}d\chi^2+%
\frac{Y_{01}Y_{03}}{4r_0r_1r_3Y_{13}}(d\rho^2+dz^2)+\frac{r_0+\zeta_0}{%
(r_1+\zeta_1)(r_3+\zeta_3)}\rho^2d\varphi^2.  \notag
\end{eqnarray}
As it is apparent from the rod structure in Figure $1a)$, in absence of black holes, one recognizes the background to be the Euclidian form of the four-dimensional C-metric with an additional trivial time direction. It is clear now that one should always expect the presence of conical defects for
any configuration of black holes in this background. In particular, as noted
previously in \cite{Tan:2003jz}, one finds unavoidable conical singularities
even in the case of the extremal charged black hole.


\subsubsection{The charged black Saturn solution}


If one chooses to have a system consisting of a black ring with a black hole
in its center (black Saturn), the appropriate harmonic function $h$ is found to be:
\begin{eqnarray}  \label{h-Saturn}
e^{2h}&=&\sqrt{\frac{(r_1+\zeta_1)(r_3+\zeta_3)(r_4+\zeta_4)}{(r_2+\zeta_2)}}%
.
\end{eqnarray}
One can easily integrate (\ref{gammap1a}) to find:
\begin{eqnarray}  \label{2gh-Saturn}
e^{2\gamma-2h}&=&\frac{1}{K_0}\left(\frac{Y_{12}Y_{23}Y_{24}}{%
r_1r_2r_3r_4Y_{13}Y_{14}Y_{34}}\right)^{\frac{1}{4}},
\end{eqnarray}
where $K_0$ is a constant to be fixed when analyzing the conical
singularities. The final solution (\ref{final5}) is again described by five
dimensionful parameters, which correspond to the masses, charges of the
black hole and black ring and the radius of the black ring. The
rod structure of this solution corresponds to Figure $1b)$. One can also consider various limits of the above solution as was performed for the double black hole solution in the previous section. In particular we checked that if one centers on the black hole horizon and sends $R\rightarrow\infty$ one obtains the metric of the single black hole. Since $R$ now describes the radius of the black ring, the other limit, in which one centers on the black ring  horizon and pushes $R\rightarrow\infty$ corresponds to making the radius of the ring very large and leads to a black string solution, as one can also infer from the rod structure of the black Saturn.

Turning now to a discussion of the conical defects, consider first the $%
ds_{\chi \rho }^{2}$ part of the metric. For the semi-infinite rod $%
z<-R/2-\sigma _{1}$ along $\chi $ one finds:
\begin{equation*}
\delta _{\chi }=2\pi -\Delta \chi \sqrt{\frac{K_{0}}{2}}.
\end{equation*}%
If we set the period of $\chi $ be $\Delta \chi =2\pi $ and choose to have a
regular outer axis (for $z<-R/2-\sigma _{1}$) we must set $K_{0}=2$.
Similarly, for the $ds_{\varphi \rho }^{2}$ part of the metric, it turns out
that conical singularities cannot be avoided and must be present either
along $-R/2+\sigma _{1}<z<R/2-\sigma _{2}$ part of the axis, where:
\begin{equation*}
\delta _{\varphi }=2\pi -\Delta \varphi \sqrt{\frac{K_{0}}{2}}\left( \frac{%
\nu +2k}{\nu -2k}\right) ^{\frac{3}{4}}\left( \frac{(R+\sigma
_{2})^{2}-\sigma _{1}^{2}}{(R-\sigma _{2})^{2}-\sigma _{1}^{2}}\right) ^{%
\frac{1}{4}},
\end{equation*}%
or along $R/2+\sigma _{2}<z$, where we find:
\begin{equation*}
\delta _{\varphi }=2\pi -\Delta \varphi \sqrt{\frac{K_{0}}{2}}.
\end{equation*}%
If one requires the outer axis $z>R/2+\sigma _{2}$ be regular, one sets the
period $\Delta \varphi =2\pi $ and $K_{0}=2$. One then finds that there
exists a conical defect in between the black ring and the black hole in its
center. It is interesting to note that in order for the charged black Saturn
system be in equilibrium one has to choose the parameters such that the
following equilibrium condition is satisfied:
\begin{equation}
\left( \frac{\nu -2k}{\nu +2k}\right) ^{3}=\left( \frac{(R+\sigma
_{2})^{2}-\sigma _{1}^{2}}{(R-\sigma _{2})^{2}-\sigma _{1}^{2}}\right) .
\label{conditie-Saturn}
\end{equation}
Notice that this relation is satisfied for any value of $R$ if one considers
configurations of extremal objects, for which $k=\sigma _{1}=\sigma _{2}=0$.
However, the resulting configuration presents a naked singularity located on
the black ring event horizon. This horizon singularity was already present in the case of a single extremally charged black ring.

We performed a numerical analysis of this equation searching for various values of the parameters describing non-extremal configurations. Although a systematic analysis of this issue is beyond the purpose of this paper, it turns out that this equation can be satisfied for families of non-extreme configurations. In the numerical analysis we fixed the length scale by taking $R=1$, and looked for solutions with positive $\sigma_i$ and positive ADM mass $i.e.$ $M_1+M_2>0$ (see the discussion in Section 3.3 below). We further imposed the condition of non-overlapping horizons $R>\sigma_1+\sigma_2$. In practice, we used a $F90$ routine to evaluate equation (\ref{conditie-Saturn}) on a equidistant grid in $(M_i,Q_i)$, for some regions of the parameters space. The $M_i,Q_i$-points satisfying (\ref{conditie-Saturn}) within some given accuracy together with the conditions mentioned above, are then used as initial values for a standard equation solver. This provides solutions of the equation (\ref{conditie-Saturn}) with an accuracy of $10^{-14}$.
Our results indicate that configurations with real $\sigma_i$ satisfying both the equilibrium and the non-overlapping horizons conditions exist. Moreover, such configurations exist for compact regions in the $(M_i,Q_i)$ parameters space. An interesting case is provided by static black Saturns in equilibrium, which exhibit zero net electric charge as measured at infinity, \textit{i.e.} $Q_1+Q_2=0$, although there is a nonvanishing local charge density.

However, in all the solutions we found, the parameters $M_1$ and $M_2$ have opposite signs, although $M_1+M_2>0$. The fact that these parameters have opposite signs does not necessarily imply that the equilibrium solutions have pathological properties. In fact, as we shall see later in Section $3.3$, the Komar masses of the individual constituents  (as computed on the horizons) are proportional not to the parameters
$M_i$ but to the parameters $\sigma_i$,  that is to the lengths of the rods determining the respective
horizons. In our case, we find that the individual Komar masses are all positive!

However, our preliminary numerical results also suggest that for all the equilibrium solutions
we found so far (with opposite signs of the mass parameters $M_i$) the denominator of $f$ (\textit{i.e.} $A+B$) seems to vanish for finite nonzero values of $(\rho,z)$ and this signals the presence of naked curvature
singularities outside the horizons since the Kretschmann scalar is proportional to $\frac{1}{(A+B)^6}$.  Thus the existence of physically relevant static charged black Saturns remains an open problem.

\subsubsection{Multi-black strings and non-extremal Majumdar-Papapetrou
solutions}


It is of interest to see the effects of ``horizon corrections'' solely,
that is, if one does not use any rod-moving factors in $h$. For this
purpose, let us consider now the following harmonic function $h$:
\begin{eqnarray}
e^{2h}&=&\sqrt{\frac{(r_1+\zeta_1)(r_3+\zeta_3)}{(r_2+\zeta_2)(r_4+\zeta_4)}}.
\end{eqnarray}
One easily integrates (\ref{gammap1a}) to find:
\begin{eqnarray}
e^{2\gamma}&=&\frac{1}{K_0}\left(\frac{16Y_{12}Y_{14}Y_{23}Y_{34}}{%
r_1r_2r_3r_4Y_{13}Y_{24}}\right)^{\frac{1}{4}},
\end{eqnarray}
where $K_0$ is a constant that can be fixed by demanding asymptotic flatness
of the solution. 
\begin{figure}[tbp]
\par
\begin{center}
\includegraphics{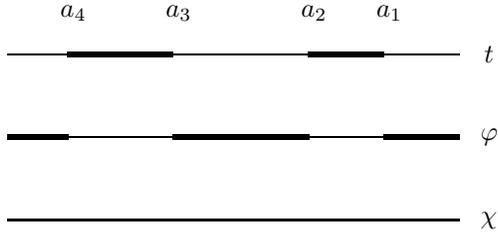} 
\end{center}
\caption{Rod structure of the double-black string system.}
\label{rod2}
\end{figure}

The rod structure of this solution is given in Figure $3$. One notices that
there is no rod along the $\chi$ direction and, therefore, our solution
should correspond to a configuration of non-extremal charged black strings.
To confirm this interpretation, let us again center on the horizon of one object and push the other to infinity. For convenience, let us center on the black hole on the right by shifting $z\rightarrow z-R/2$, and then taking the limit $R\rightarrow\infty$. We find that Manko's solution (\ref{Manko}) reduces to the single Reissner-Nordstr\"om black hole solution, while:
\begin{eqnarray}
e^{2h}&=&\sqrt{\frac{r_{1}+\zeta_1}{r_{2}+\zeta_2}},~~~~~~~ e^{2\gamma}=%
\frac{1}{K_0}\left(\frac{16Y_{12}} {r_{1}r_{2}}\right)^{\frac{1}{4}}.
\end{eqnarray}
Performing now the coordinate transformations:
\begin{eqnarray}
\rho&=&\sqrt{(r-m)^2-\sigma^2}\sin\theta,~~~~~~~z=(r-m)\cos\theta,
\end{eqnarray}
and by appropriately choosing the value of $K_0$ one obtains the uniform black string solution:
\begin{eqnarray}
ds^2&=&-\frac{(r-m)^2-\sigma^2}{r^2}dt^2+\frac{rdr^2}{r-m-\sigma}+\frac{%
rd\chi^2}{r-m+\sigma}+r(r-m+\sigma)(d\theta^2+\sin^2\theta d\varphi^2),
\notag \\
A_t&=&\frac{\sqrt{3(m^2-q^2)}}{r},
\end{eqnarray}
as advertised.

Turning now to the discussion of the conical singularities, one finds that there is a conical singularity:
\begin{eqnarray}
\delta_{\varphi}&=&2\pi-\Delta\varphi\sqrt{\frac{K_0}{\sqrt{8}}}.
\end{eqnarray}
along the outer axis $z<-R/2-\sigma_1$ or $z>R/2+\sigma_2$, while:
\begin{eqnarray}
\delta_{\varphi}&=&2\pi-\Delta\varphi\sqrt{\frac{K_0}{\sqrt{8}}}\left(\left(%
\frac{\nu+2k}{\nu-2k}\right)^{3}\frac{R^2-(\sigma_1+\sigma_2)^2}{%
R^2-(\sigma_1-\sigma_2)^2}\right)^{\frac{1}{4}},
\end{eqnarray}
on the portion $-R/2+\sigma_1<z<R/2-\sigma_2$ in between the black string
horizons. We ensure regularity of the outer axis, by taking $%
\Delta\varphi=2\pi$ and setting $K_0=\sqrt{8}$. There will still be a
conical singularity running in between the black strings. The equilibrium
condition, for which this conical singularity disappears is given by:
\begin{eqnarray}  \label{conditie-UBS}
\left(\frac{\nu-2k}{\nu+2k}\right)^3&=&\frac{R^2-(\sigma_1+\sigma_2)^2}{%
R^2-(\sigma_1-\sigma_2)^2}.
\end{eqnarray}
However, when solving this equation numerically, we failed to find
nonextremal solutions with $\delta_{\varphi}=0$ also satisfying the physical
conditions $M_1+M_2>0$ and $\sigma_1+\sigma_2<R$.\footnote{The numerical methods here were similar to those employed for the black Saturn case.}

One clear way to satisfy (\ref{conditie-UBS}) is to consider extremal
objects for which $M_1=Q_1$ and $M_2=Q_2$. Again, this leads to $%
\sigma_1=\sigma_2=k=\mu=0$ and, in consequence, $r_1=r_2$ and $r_3=r_4$.
Using (\ref{extremal4dManko}) and noticing that in the extremal
limit one has $h=0$, and therefore from (\ref{gammap1a}) $\gamma=0$, the final metric for the
extremal double black string solution takes the simple form:
\begin{eqnarray}
ds_{MP}^2&=&-\left(1+\frac{M_1}{r_3}+\frac{M_2}{r_1}\right)^{-2}dt^2+\left(1+%
\frac{M_1}{r_3}+\frac{M_2}{r_1}\right)\big[d\chi^2+d\rho^2+dz^2+\rho^2d%
\varphi^2\big],  \notag \\
A_t&=&-\frac{\sqrt{3}}{2}\left(1+\frac{M_1}{r_3}+\frac{M_2}{r_1}\right)^{-1}.
\end{eqnarray}
The metric inside the square bracket describes Euclidian flat space and
one recognizes the above solution as the particular case of the extremal
Majumdar-Papapetrou double-black hole solution \cite{Myers:1986rx,
Duff:1993ye}.


\subsection{Basic properties of the new solutions}


All relevant quantities of the five-dimensional solutions can be expressed
in terms of the parameters $M_i,Q_i$ and $R$, which enter the
four-dimensional seed solution. Explicitly, one finds that the Hawking
temperature and event horizon area receive corrections that are fixed
by the explicit form of the harmonic function $h$. Note that in the
five-dimensional solution the horizons are still located at $\rho=0$, $%
a_2<z<a_1$ (upper black object) and $\rho=0$, $a_4<z<a_2$ (lower black
object).

Near the black hole horizons, the leading order expressions of the functions
that appear in the general line element (\ref{rel1}) are given by:
\begin{eqnarray}
e^{2h}&\sim &\sqrt{\rho}, ~~~~~~~e^{\tilde{\mu}}\sim {\rho},~~~~~~~\tilde
f\sim {\rho^2},
\end{eqnarray}
where the proportionality factors depend on $z$. Note that the
following relations also hold near horizons, as implied by the $(\rho,z)$%
-component of Einstein's equations:
\begin{eqnarray}  \label{def-p}
e^{4h-4\gamma}= (p^{(i)})^4 \rho+O(\rho^2).
\end{eqnarray}
The constant $p^{(i)}$, for each $i=1,2$ is fixed by the expression of $h$
and takes different values for the upper and lower horizons.

The Hawking temperatures for each constituent of the five-dimensional
solution can be computed either by evaluating the surface gravity or from the Euclidian section and it can be expressed as:
\begin{equation}
T_{H}^{(i)}=\frac{(\kappa ^{(i)})^{3/4}p^{(i)}}{2\pi },
\label{temperatura}
\end{equation}%
where $\kappa ^{(i)}$ is the surface gravity of the $i^{th}$-black hole for
the four-dimensional seed metric (\ref{Manko}). The constants $\kappa ^{(i)}$
are given by the relation (\ref{kappa}) in the Appendix, where they are
expressed in terms of $M_{1}$, $Q_{1}$, $M_{2}$, $Q_{2}$ and $R$.

The horizon area of the $i^{th}$-black object is given by:
\begin{eqnarray}  \label{area5d}
A_h^{(i)}=\frac{4\pi^2 }{(\kappa^{(i)})^{3/4}p^{(i)}}\Delta z^{(i)},
\end{eqnarray}
for $i=1..2$, where $\Delta z^{(1)}=a_1-a_2=2\sigma_2$, $\Delta
z^{(2)}=a_3-a_4=2\sigma_1$. As usual, one identifies the entropy with one
quarter of the event horizon area.

By using (\ref{rel1}) and the relations (\ref{V4d-sim}) in the Appendix, it
is strainghtforward to show that the electric potential on the horizon of
one of the black holes is:
\begin{equation}
\Phi ^{(i)}=\frac{\sqrt{3}}{2}\left( \frac{M_{i}-\sigma _{i}}{Q_{i}}\right) ,
\end{equation}%
while the electric charges are evaluated according to:\footnote{%
Here the integration is performed at the horizon.}
\begin{equation}
Q_{e}^{(i)}=\frac{1}{8\pi G }\int_{S}F_{\mu \nu}dS^{\mu\nu},
\end{equation}%
and one obtains $Q_{e}^{(i)}=\frac{\sqrt{3}\pi}{G} Q_{i}$. If instead one computes 
this integral on the three-sphere at infinity enclosing both the black holes
one finds the total electric charge $Q_{e}=Q_{e}^{(1)}+Q_{e}^{(2)}$.

To compute the ADM mass of the solutions, one performs the coordinate change
$\rho=\frac{1}{2}r^2\sin 2\theta$, $z=\frac{1}{2}r^2\cos 2\theta$, and
evaluates the expression of $g_{tt}$ as $r\to \infty$. One finds that
the total mass is given by:
\begin{eqnarray}
M_{ADM}=\frac{3\pi}{2G}(M_1+M_2).
\end{eqnarray}

One can also evaluate the Komar mass of an individual black object, by using
the definition:
\begin{equation}
M=-\frac{1}{16\pi G }\frac{3}{2}\int_{S}\alpha \ ,
\label{MK}
\end{equation}
where $S$ is the boundary of any spacelike hypersurface and:
\begin{equation}
\alpha _{\mu \nu \rho }=\epsilon _{\mu \nu \rho \sigma \tau }\nabla ^{\sigma
}\xi ^{\tau }\ ,
\end{equation}%
with the Killing vector $\xi =\partial /\partial t$. This relation measures
the mass contained in $S$, and therefore the horizon mass $M_{H}$ is
obtained by performing the above integration at the horizon. If we
take $S$ to be the three-sphere at infinity enclosing both horizons instead, then (\ref{MK}) gives the total mass of the system, which coincides with the ADM mass. A straightforward computation leads to:
\begin{equation}
M_{Komar}^{(1)}=\frac{3}{8\pi G}\sigma _{2}\Delta \varphi \Delta \chi
,~~~~~~~M_{Komar}^{(2)}=\frac{3}{8\pi G}\sigma _{1}\Delta \varphi \Delta \chi
,
\end{equation}%
while
\begin{equation}
M=M_{Komar}^{(1)}+M_{Komar}^{(2)}-\frac{1}{16\pi G}\frac{3}{2}\int R_{t}^{t}%
\sqrt{-g}dV.
\end{equation}%
However, since Einstein's equations imply $R_{t}^{t}=\frac{F_{\mu t}^{2}%
}{3}$, one arrives at the following five-dimensional Smarr formula \cite%
{Gibbons:1987ps}:
\begin{equation*}
M=\mathcal{M}^{(1)}+\mathcal{M}^{(2)},
\end{equation*}%
where for each constituent one has:\footnote{This relation follows from the four-dimensional Smarr formula (\ref{Smarr-4D})}
\begin{equation*}
\frac{2}{3}\mathcal{M}^{(i)}=T_{H}^{(i)}S^{(i)}+\frac{2}{3}\Phi
_{H}^{(i)}Q_{e}^{(i)}~,
\end{equation*}%
where $\mathcal{M}^{(i)}=\frac{3\pi }{2G}M_{i}$. Thus one can regard $%
\mathcal{M}^{(i)}$ as the individual mass of each black object, containing
an electromagnetic contribution apart from the Komar part! Notice that there
is no compelling reason to impose $\mathcal{M}^{(i)}>0$ as long as the total
mass $M$ measured at infinity is still positive. Moreover, the Komar mass of each
constituent is positive, since it is proportional to the length of the rod determining each horizon. Finally, let us notice that for $|Q_i|=M_i$ the extremality condition is indeed satisfied \cite{Gauntlett:1998fz}:
\beqs
\frac{{\cal M}^{(i)}}{|Q_e^{(i)}|}&=&\frac{\sqrt{3}}{2}.
\eeqs

The situation is slightly different for the double black string solutions.
Since the background approached at infinity  is ${\cal M}^4\times S^1$,
the black strings have also a nonzero tension, which is the charge associated 
with the Killing vector $\partial/\partial \chi$. To find the ADM mass and tension of 
 the uniform black strings, we consider the asymptotics $g_{tt},~g_{\chi\chi}$
 in a coordinate system with $\rho=r \sin \theta$, $z=r \cos \theta$: 
\beqs
\label{UBS-asympt}
g_{tt}=-1+\frac{c_t}{r}+O(1/r^2),~~g_{\chi\chi}=1+\frac{c_\chi}{r}+O(1/r^2),
\eeqs
with $c_t=2(M_1+M_2)$, $c_\chi=M_1+M_2-\sigma_1-\sigma_2$.
Thus \cite{Harmark:2007md}
\beqs
\label{UBS-MT}
M_{ADM}=\frac{LV_2}{16\pi G}(2c_t-c_\chi),~~{\cal ~T}_{ADM}=\frac{V_2}{16\pi G}( c_t-2c_\chi),
\eeqs
where $V_2=4 \pi$ and $L=\Delta\chi$,
which agree with the Komar mass and tension $M$, ${\cal T}$.
The black strings satisfy the Smarr relation: 
\beqs
\label{UBS-Smarr}
\frac{1}{3}(2M-L {\cal T})=T_H^{(1)} S^{(1)}+T_H^{(2)} S^{(2)}+\frac{2}{3}(\Phi_H^{(1)}Q_e^{(1)}+\Phi_H^{(2)}Q_e^{(2)} ).
\eeqs

The above relations allow a discussion of the basic physical properties of
the solutions we found. Take for instance the double black hole solution
discussed in Section $3.2.1$. The first black hole horizon is located at $%
\rho=0$ for $a_2\leq z\leq a_1$ and the metric on a spatial cross-section of
the horizon can be written as:
\begin{eqnarray}  \label{metric-horizon-1}
&ds^2_{BH^1}=\sqrt{\frac{\zeta_2\zeta_3\zeta_4}{|\zeta_1|}}\frac{1}{z\sqrt{%
F^{(1)}(z) }}d\chi^2 +\frac{F^{(1)}(z) }{(\kappa^{(1)})^{3/2}(p^{(1)})^2}%
dz^2 +\sqrt{\frac{|\zeta_1|}{\zeta_2\zeta_3\zeta_4}}\frac{z}{\sqrt{%
F^{(1)}(z) }}d\varphi^2,
\end{eqnarray}
where, as implied by (\ref{def-p}):
\begin{eqnarray}
p^{(1)} =\left(\frac{2K_0^2\sigma_2(R-\sigma_1+\sigma_2)(R+\sigma_1+\sigma_2)%
}{(R+2\sigma_2)^2} \right)^{1/4}.
\end{eqnarray}
The second black hole horizon is located at $\rho=0$ for $a_4\leq z\leq a_3$%
. The metric on a spatial cross-section of the horizon is:
\begin{eqnarray}  \label{metric-horizon-2}
&ds^2_{BH^2}=\sqrt{\frac{ \zeta_4}{|\zeta_1\zeta_2\zeta_3|}}\frac{z}{\sqrt{%
F^{(2)}(z) }}d\chi^2 +\frac{F^{(2)}(z) }{(\kappa^{(2)})^{3/2}(p^{(2)})^2}%
dz^2 +\frac{1}{z}\sqrt{\frac{|\zeta_1\zeta_2\zeta_3|}{\zeta_4}}\frac{1}{%
\sqrt{F^{(2)}(z) }}d\varphi^2,
\end{eqnarray}
where
\begin{eqnarray}
p^{(2)} =\left( \frac{2K_0^2\sigma_2(R+\sigma_1-\sigma_2)(R+\sigma_1+%
\sigma_2)}{(R+2\sigma_1)^2} \right)^{1/4}.
\end{eqnarray}
One can easily see that the topology of the horizon is $S^3$ in both cases,
as expected from the rod diagram in Figure $1a)$.

Turning now to the charged black Saturn solution derived in Section $3.2.2$,
the black ring horizon is located at $\rho=0$ for $a_2\leq z\leq a_1$. The
metric of the spatial cross section of the black ring horizon reads:
\begin{eqnarray}  \label{metric-horizon-ring}
&ds^2_{BR}=\sqrt{\frac{\zeta_3\zeta_4}{|\zeta_1|\zeta_2}}\frac{1}{\sqrt{%
F^{(1)}(z) }}d\chi^2 +\frac{F^{(1)}(z) }{(\kappa^{(1)})^{3/2}(p^{(1)})^2}%
dz^2 +\sqrt{\frac{|\zeta_1|\zeta_2}{\zeta_3\zeta_4}}\frac{1}{\sqrt{%
F^{(1)}(z) }}d\varphi^2,
\end{eqnarray}
where we denoted:
\begin{eqnarray}
p^{(1)} =\left(\frac{K_0^2 (R+\sigma_2-\sigma_1)(R+\sigma_1+\sigma_2)}{4
\sigma_2}\right)^{1/4}.
\end{eqnarray}
Along the black ring horizon, the orbits of $\varphi$ shrink to zero at $%
z=a_1$ and $z=a_2$, while the orbits of $\chi$ do not shrink to zero
anywhere. Thus the topology of the horizon is $S^2\times S^1$ as expected
from the rod diagram in Figure $1b)$. However, the black hole
horizon is located at $\rho=0$ for $a_4\leq z\leq a_3$. The metric on the
spatial cross-section of the black hole horizon is in this case given by:
\begin{eqnarray}  \label{metric-horizon-hole}
&ds^2_{BH}=\sqrt{\frac{|\zeta_2|\zeta_4}{|\zeta_1\zeta_3|}}\frac{1}{\sqrt{%
F^{(2)}(z) }}d\chi^2 +\frac{F^{(2)}(z) }{(\kappa^{(2)})^{3/2}(p^{(2)})^2}%
dz^2 +\sqrt{\frac{|\zeta_1\zeta_3|}{|\zeta_2|\zeta_4}}\frac{1}{\sqrt{%
F^{(2)}(z) }}d\varphi^2,
\end{eqnarray}
where:
\begin{eqnarray}
p^{(2)} =\left(\frac{ K_0^2\sigma_1(R+\sigma_1+\sigma_2) }{%
(R+\sigma_1-\sigma_2)}\right)^{1/4}.
\end{eqnarray}
The orbits of $\varphi$ shrink to zero at $z=a_3$, while the orbits of $\chi$
shrink to zero at $z=a_4$. Thus the topology of the horizon is indeed $S^3$
and it corresponds to a black hole.

A similar computation can be performed for the two-black string solution
presented in Section 3.2.3. The metric of the spatial cross section of black
strings horizons reads:
\begin{eqnarray}  \label{metric-horizon-UBS1}
&&ds^2_{BS_1}=\sqrt{\frac{\zeta_3 }{|\zeta_1|\zeta_2\zeta_4}}\frac{1}{2\sqrt{%
F^{(1)}(z) }}d\chi^2+\frac{F^{(1)}(z) }{(\kappa^{(1)})^{3/2}(p^{(1)})^2}dz^2
+\sqrt{\frac{|\zeta_1|\zeta_2\zeta_4 }{\zeta_3 }}\frac{2}{\sqrt{F^{(1)}(z) }}%
d\varphi^2,  \label{metric-horizon-UBS1}
\end{eqnarray}
for the first black string horizon, respectively:
\begin{eqnarray}
ds^2_{BS_2}&=&\sqrt{\frac{|\zeta_2| }{ \zeta_1 \zeta_3\zeta_4}}\frac{1}{2%
\sqrt{F^{(2)}(z) }}d\chi^2+\frac{F^{(2)}(z) }{(\kappa^{(1)})^{3/2}(p^{(2)})^2%
}dz^2 +\sqrt{\frac{\zeta_1 \zeta_3\zeta_4 }{|\zeta_2|}}\frac{2}{\sqrt{%
F^{(2)}(z) }}d\varphi^2,
\end{eqnarray}
for the second black string horizon, where we denoted:
\begin{eqnarray}
p^{(1)} =\left(\frac{K_0^2 (R+\sigma_2-\sigma_1)}{32 \sigma_2
(R+\sigma_1+\sigma_2)}\right)^{1/4},~~ p^{(2)} = \left(\frac{K_0^2
(R+\sigma_1-\sigma_2)}{32 \sigma_1 (R+\sigma_1+\sigma_2)}\right)^{1/4}.
\end{eqnarray}
One can see that the orbits of $\varphi$ shrink to zero at $z=a_1$ and $%
z=a_2 $ for the first black string, and at $z=a_3$ and $z=a_4$ for the
second black string, while the orbits of $\chi$ do not shrink to zero
anywhere. Thus the topology of the horizon is indeed $S^2\times S^1$ as
expected from the rod diagram in Figure $3$.


\section{Conclusions}

In this paper, by using a novel solution generation technique we were able to
construct the general non-extremally charged multi-black hole solutions in five dimensions.
As opposed to other solution-generating methods used previously in the literature to construct
five-dimensional solutions \cite{Teo:2003ug,Yazadjiev:2007cd,Yazadjiev:2006hw,Azuma:2007qj} our method lifts a four-dimensional static charged solution of Einstein-Maxwell field equations to a solution in the more general Einstein-Maxwell-Dilaton theory in five dimensions. While the fields of the general EMD solution can be read in each case from (\ref{final5}), in discussing the generated solutions we focused for simplicity on Einstein-Maxwell theory, for which the coupling constant $\alpha=0$ in the general solution (\ref{final5}) vanishes.

For simplicity we restricted our attention to configurations
consisting of only two constituents. In four dimensions there exists a
general solution describing a static configuration of two
Reissner-Nordstr\"om black holes, which was recently cast into a simpler
form in \cite{Manko:2007hi,Alekseev:2007re}. When lifted to five dimensions,
we found solutions describing general static configurations of charged black
objects. These solutions include as particular cases the charged black
Saturn, the double non-extremal Reissner-Nordstr\"om solution and the double
black string solution, whose extremal limit we found to be precisely of the
five dimensional Majumdar-Papapetrou type solution. Even though we found static configurations of non-extremal black holes/rings/strings, our numerical results indicate the presence of naked curvature singularities for finite values of $\rho$ (outside the horizons at $\rho = 0$) for all charged black Saturn solutions satisfying the equilibrium condition.  We are skeptical that any equilibrium charged black saturn solutions exist without angular momenta.

We also note that for suitable choices of the harmonic function $h$
the generated solutions can describe configurations of two black rings
(orthogonal or concentric). These double-ring solutions would correspond to
the charged versions of the static di-ring solution and the bicycling black
ring system \cite{Iguchi:2007is,Mishima:2005id,Iguchi:2006rd,Iguchi:2006tu,Izumi:2007qx,Elvang:2007hs,Evslin:2007fv}. For example, in the di-ring
case one takes:
\begin{eqnarray}
e^{2h}&=&(r-\zeta_0)\sqrt{\frac{(r_1+\zeta_1)(r_3+\zeta_3)}{%
(r_2+\zeta_2)(r_4+\zeta_4)}},
\end{eqnarray}
where $\zeta_0=z-\sigma_0$, with $\sigma_0>\sigma_2$, while for the
charged bi-ring system one takes:
\begin{eqnarray}
e^{2h}&=&(r+\zeta_0)\sqrt{\frac{(r_1+\zeta_1)(r_4+\zeta_4)}{%
(r_2+\zeta_2)(r_3+\zeta_3)}},
\end{eqnarray}
where $\zeta_0=z$. In both cases it is trivial to integrate (\ref%
{gammap1a}) to find explicitly the factor $e^{2\gamma}$, which enters the
general solution.

One should also remark at this point that instead of using the
four-dimensional double-Reissner Nordstr\"om as the seed metric, we could have used the more general solution describing general configurations of $N$ Reissner-Nordstr\"om black holes given in \cite{Ruiz:1995uh}. For such a configuration the choice of the harmonic function $h$ in our solution-generating technique is easily inferred from the rod diagram of the
solution one wishes to describe. One also has to add the correction factor for each black object horizon as described in Section $2$.

As a general remark, since all the five dimensional solutions constructed
using the double-Reissner-Nordstr\"om four dimensional solution as the seed metric are sufficiently complicated, it is generally a very difficult task to check algebraically that they satisfy the Einstein-Maxwell equations. However, the correctness of our general solutions was confirmed via numerical methods and we explicitly verified that our solutions satisfy the field equations for several sets of constants $(M_i,Q_i,R)$.

\vspace{10pt}

{\Large Acknowledgements}

The work of R.B.M. and C.S. was supported by the Natural Sciences and
Engineering Research Council of Canada.

\renewcommand{\theequation}{A-\arabic{equation}}
\setcounter{equation}{0} 


\section*{A: The double-Reissner-Nordstr\"om solution in four dimensions}

We briefly present the basic properties of the double-Reissner-Nordstr\"om solution in four dimensions.
In the parameterization given recently by Manko in \cite{Manko:2007hi}, the four-dimensional quantities that appear in (\ref{Manko})
are:
\begin{eqnarray}  \label{Mankonew}
A=\sum_{1\leq i<j\leq 4}a_{ij}r_i r_j,~~~B=\sum_{i=1}^4 b_i r_i,
~~~C=\sum_{i=1}^4 c_i r_i,
\end{eqnarray}
where we defined:
\begin{eqnarray}
&&a_{12}=a_{34}=4k \sigma_1\sigma_2,~a_{13}= a_{24}=2k^2-\nu(\mu^2
-\sigma_1\sigma_2),~~a_{14}= a_{23}=-2k^2+\nu(\mu^2 +\sigma_1\sigma_2),
\notag \\
&&b_1=2\sigma_1 [\nu(-\mu(\mu+ Q_2)+M_1\sigma_2) +2k (\mu (\mu-
Q_1)+M_2(\sigma_2 -R)) ],  \notag \\
&&b_2=2\sigma_1 [\nu(\mu(\mu+ Q_2)+M_1\sigma_2) +2k (\mu (-\mu+
Q_1)+M_2(\sigma_2+R)) ],  \notag \\
&&b_3=2\sigma_2 [\nu(\mu(\mu- Q_1)+M_2\sigma_1) -2k (\mu (\mu+
Q_2)-M_1(\sigma_1+R)) ],  \notag \\
&&b_4=2\sigma_2 [\nu(\mu(-\mu+ Q_1)+M_2\sigma_1) +2k (\mu (\mu+
Q_2)+M_1(\sigma_1-R)) ],  \notag \\
&&c_1=-2\sigma_1 [2k (\mu M_1+(\mu
+Q_2)(R-\sigma_2))+\nu(M_2\mu+\sigma_2(\mu-Q_1)) ],  \notag \\
&&c_2=2\sigma_1 [2k (\mu M_1+(\mu
+Q_2)(R+\sigma_2))+\nu(M_2\mu+\sigma_2(-\mu+Q_1)) ],  \notag \\
&&c_3=-2\sigma_2 [2k (\mu
M_2+(\mu-Q_1)(R+\sigma_1))+\nu(M_1\mu-\sigma_1(\mu+Q_2)) ],  \notag \\
&&c_4=2\sigma_2 [2k (\mu
M_2+(\mu-Q_1)(R-\sigma_1))+\nu(M_1\mu+\sigma_1(\mu+Q_2)) ].  \label{coeff}
\end{eqnarray}
Depending on the values of $M_i,Q_i,R$, this solution describes two
non-extremal black holes, two naked singularities or a black hole-naked
singularity. The equlibrium is possible only in this last case.
However, we shall consider only the case of real $\sigma_i$,
corresponding to a configuration of two interacting black holes. The black
hole event horizons are located at $\rho=0$ and $a_2<z<a_1$ (the first black
hole) and $a_4<z<a_3$ (the second black hole).

It is also useful to present the approximate expressions of the basic pieces
$A,B,C$ near one of the horizons:
\begin{eqnarray}
A(\rho ,z) &=&A_{0}(z)+A_{2}(z)\rho ^{2}+O(\rho ^{4}),~~\mathrm{with}%
~~A_{0}=\sum_{1\leq i<j\leq 4}a_{ij}|\zeta _{i}\zeta _{j}|,~~A_{2}=\frac{1}{2%
}\sum_{1\leq i<j\leq 4}a_{ij}\frac{(\zeta _{i}^{2}+\zeta _{j}^{2})}{|\zeta
_{i}\zeta _{j}|},  \notag  \label{A} \\
B(\rho ,z) &=&B_{0}(z)+B_{2}(z)\rho ^{2}+O(\rho ^{4}),~~\mathrm{with}%
~~B_{0}=\sum_{i=1}^{4}b_{i}|\zeta _{i}|,~~B_{2}=\frac{1}{2}%
\sum_{i=1}^{4}b_{i}\frac{1}{|\zeta _{i}|},  \notag \\
C(\rho ,z) &=&C_{0}(z)+C_{2}(z)\rho ^{2}+O(\rho ^{4}),~~\mathrm{with}%
~~C_{0}=\sum_{i=1}^{4}c_{i}|\zeta _{i}|,~~C_{2}=\frac{1}{2}%
\sum_{i=1}^{4}c_{i}\frac{1}{|\zeta _{i}|},
\end{eqnarray}%
which implies the following near horizon expression for the function $f$:
\begin{equation*}
f(\rho ,z)=F(z)\rho ^{2}+O(\rho ^{4}),~~~\mathrm{with}~~~~F(z)=2\frac{%
A_{0}A_{2}-B_{0}B_{2}+C_{0}C_{2}}{(A_{0}+B_{0})^{2}}.
\end{equation*}%
We emphasize that $F(z)$ has a different
expression for each horizon, the corresponding function being labeled as $%
F^{(i)}(z)$, $i=1,2$. One can easily see that as $z\rightarrow a_{i}$ (near the ends of the rods), $F(z)\sim 1/|\zeta _{i}|$.

The two horizons have different Hawking temperatures, which are given by:
\begin{eqnarray}  \label{TH4d}
T_H^{(i)}=\frac{\kappa^{(i)}}{2\pi}
\end{eqnarray}
with the surface gravities:
\begin{eqnarray}  \label{kappa}
\kappa^{(i)}=\frac{k_0^{(i)}}{t_0^{(i)}}\sqrt{t_A^{(i)}-t_B^{(i)}+t_C^{(i)}}
\end{eqnarray}
where:
\begin{eqnarray}  \label{expr1}
&&t_A^{(k)}=-a_1a_2a_3a_4\bigg(\sum_{1\leq i<j\leq 4}a_{ij}a_ia_i
\epsilon_{ij}^{(k)} \bigg)\bigg(\sum_{1\leq l<m\leq
4}a_{lm}\epsilon_{lm}^{(k)}\frac{(a_l^2+a_m^2)}{a_la_m }\bigg),  \notag \\
&&t_B^{(k)}=-a_1a_2a_3a_4(\sum_{i=1}^4\frac{b_i}{a_i}\epsilon_{ii}^{(k)})(%
\sum_{j=1}^4 a_j b_j\epsilon_{jj}^{(k)}), \\
&& t_C^{(k)}=-a_1a_2a_3a_4(\sum_{i=1}^4\frac{c_i}{a_i}\epsilon_{ii}^{(k)})
(\sum_{j=1}^4 c_j b_j\epsilon_{jj}^{(k)}), ~~~~~~~t_0^{(k)}=(\sum_{1\leq
i<j\leq 4}a_{ij}a_ia_j\epsilon_{ij}^{(k)}+\sum_{i=1}^4 a_i
b_i\epsilon_{ii}^{(k)})^2.  \notag
\end{eqnarray}
The symbol $\epsilon_{ij}^{(k)}$ is defined such that for the upper black
hole one takes:
\begin{eqnarray}  \label{epsilon}
-\epsilon_{11}^{(1)}=\epsilon_{22}^{(1)}=\epsilon_{33}^{(1)}=%
\epsilon_{44}^{(1)}=1,~~
\epsilon_{12}^{(1)}=\epsilon_{13}^{(1)}=\epsilon_{14}^{(1)}=-%
\epsilon_{23}^{(1)}=-\epsilon_{24}^{(1)}=-\epsilon_{34}^{(1)}=1,
\end{eqnarray}
while for the lower black hole one takes:
\begin{eqnarray}  \label{epsilon}
\epsilon_{11}^{(2)}=\epsilon_{22}^{(2)}=\epsilon_{33}^{(2)}=-%
\epsilon_{44}^{(2)}=1,~~
\epsilon_{12}^{(2)}=\epsilon_{13}^{(2)}=\epsilon_{23}^{(2)}=-%
\epsilon_{14}^{(2)}=-\epsilon_{24}^{(2)}=-\epsilon_{34}^{(2)}=1.
\end{eqnarray}
The event horizon area of each black hole is given by:
\begin{eqnarray}  \label{area4d}
A_h^{(i)}=\frac{2\pi }{\kappa^{(i)}}\Delta z^{(i)},
\end{eqnarray}
where we denoted $\Delta z^{(1)}=a_1-a_2=2\sigma_1$ and $\Delta
z^{(2)}=a_3-a_4=2\sigma_2$.

The electric charges of the black holes are $Q_1$ for the upper one and $Q_2$
for the lower black hole \cite{Manko:2007hi}. The electrostatic potential on
the black hole horizons is constant and can be generally expressed as:
\begin{eqnarray}  \label{V4d}
V^{(k)}=- \frac{2\sum_{i=1}^4 a_i c_i\epsilon_{ii}^{(k)} }{ \sum_{1\leq
i<j\leq 4}a_{ij}a_ia_j\epsilon_{ij}^{(k)} +\sum_{i=1}^4 a_i
b_i\epsilon_{ii}^{(k)}}
\end{eqnarray}
Unfortunately, after replacing $a_{ij},a_i,b_i,c_i$ in the expressions of $%
T_H^{(i)},A_h^{(i)}$ the resulting formulae in terms of $M_i,Q_i,R$ cannot be further simplifed. However, $V^{(k)}$ can be
re-expressed in the simple form:
\begin{eqnarray}  \label{V4d-sim}
V^{(i)}=\frac{M_i- \sigma_i }{Q_i}.
\end{eqnarray}
The total ADM mass of the system is $M=M_1+M_2$. The following Smarr relation also holds in four-dimensions \cite{Gibbons:1987ps}:
\begin{eqnarray}  \label{Smarr-4D}
M_i=\frac{1}{2}T_H^{(i)}A_h^{(i)}+ Q_iV^{(i)}.
\end{eqnarray}


\end{document}